# Linear pattern matching on sparse suffix trees

Roman Kolpakov*     Gregory Kucherov†

Tatiana Starikovskaya‡

**Abstract**

Packing several characters into one computer word is a simple and natural way to compress the representation of a string and to speed up its processing. Exploiting this idea, we propose an index for a packed string, based on a *sparse suffix tree* [8] with appropriately defined suffix links. Assuming, under the standard unit-cost RAM model, that a word can store up to $\log_\sigma n$ characters ($\sigma$ the alphabet size), our index takes $O(n/\log_\sigma n)$ space, i.e. the same space as the packed string itself. The resulting pattern matching algorithm runs in time $O(m + r^2 + r \cdot occ)$, where $m$ is the length of the pattern, $r$ is the actual number of characters stored in a word and $occ$ is the number of pattern occurrences.

## 1 Introduction

Many application areas, such as genomics or computer security for example, face a sharp growth of volumes of available data. Even with the spectacular development of hardware capacities, data size often remains a bottleneck for its efficient processing, which requires new algorithmic solutions allowing for both a compact representation and efficient querying of data.

With this motivation, research in combinatorial pattern matching recently developed different sophisticated methods for efficient compact representation sequence data, see the survey [9] and references therein. A basic goal of these methods is to index a text (sequence) through a *succinct data structure*, i.e. one taking $O(n \log \sigma)$ bits of memory (as opposed to $O(n \log n)$ bits for classical indexes), where $n$ is the text length and $\sigma$ the alphabet size. Thus, succinct data structures take asymptotically as muchs pace as the text itself. Still, these indexes can efficiently support queries on the text, and primarily the string matching operation. As a downside, many of these methods, while being mathematically elegant and highly nontrivial, are too complex to be used in practice. However, some of them gave rise to practical implementation and real-life applications (e.g. [6, 11]).

---

*Moscow University, Russia. E-mail `foroman@mail.ru`

†CNRS, Laboratoire d'Informatique Gaspard-Monge, Université Paris-Est, Marne-la-Vallée, France. E-mail `Gregory.Kucherov@univ-mlv.fr`

‡Moscow University, Russia. E-mail `tat.starikovskaya@gmail.com`

One simple idea for saving memory for storing sequence data is to pack several characters into one machine word. Under the standard unit-cost RAM computation model, it is assumed that the machine word size $w$ is at least $\log n$ bits, and therefore can store as many as $\frac{\log n}{\log \sigma} = \log_\sigma n$ characters. Not only this saves memory, but also allows one to speed up algorithms, as two tuples of characters each stored in one word can then be compared with unit cost. For example, [2] recently proposed a modification of the Knuth-Morris-Pratt (KMP) pattern matching algorithm for packed strings reporting in $O(n/\log_\sigma n + m + occ)$ time all $occ$ occurrences of a pattern of size $m$. Like in the regular KMP algorithm, the text is not pre-processed and the speed-up is achieved by designing a special data structure representing the pattern.

In this paper, our goal is complementary: based on the character packing, we want to propose an index data structure for a text which supports pattern matching queries in time linear in $m$, and at the same time uses $O(n/\log_\sigma n)$ space (i.e. $O(n \log \sigma)$ bits), that is constitutes a succinct index.

The central idea of defining the index is to partition the text into blocks of $r$ characters and to construct a suffix tree which stores only those suffixes that start at the block boundaries. Such a suffix tree, called *an evenly spaced sparse suffix tree*, has been first studied in [8] (see also [1]). A suffix tree we use here differs from that of [8] in the definition of suffix links.

A sparse suffix tree allows one to easily search for pattern's occurrences that start at block boundaries. Therefore, the pattern matching procedure splits up into two parts: locating occurrences of the pattern's suffixes $P[k+1..m]$, for $k = 0..r-1$, at block boundaries and then selecting from them those locations which are preceded by the corresponding pattern prefix $P[1..k]$. To solve the second task, we use another data structure: the compacted trie of reversed blocks augmented by additional arrays assigned to its nodes. Selecting all positions corresponding to a given suffix amounts to traversing the trie and recomputing the interval of lexicographically ordered suffixes (see details in Section 5). This is done using a technique inspired from *fractional cascading* [10], which is also closely related to *wavelet trees* [5], a popular technique in text compression and indexing (see [9]). As a result, we obtain a pattern matching algorithm working in time $O(m + r^2 + r \cdot occ)$ while using space $O(n/\log_\sigma n)$ for storing both the text in packed form and working data structures.

As far as related papers are concerned, similar ideas appear in papers [3, 7, 4], although the idea of "character packing" is somewhat implicit there. Compared to those papers, our approach is different in several aspects. First, we use a sparse suffix tree over the alphabet of characters, rather than a suffix tree over the meta-alphabet of $r$-tuples of characters. Instead of searching for $r$ suffixes of the pattern independently, we locate them in a single traversal of the suffix tree, using appropriately defined suffix links. Second, we don't make use of external *orthogonal range queries* algorithms (see [3, 7]), but instead use specially designed algorithms on "classic" data structures (compacted trie). Moreover, we restrict the use of the RAM model to manipulating packed strings (i.e. to unit-cost operations on several letters packed into one machine word) and indexing packed strings following the Four-Russians idea. However, we don't use special data structures (such as the *Geometric Burrows-Wheeler transform* [3]) involving numeric computations. Overall, we obtain a fully linear pattern matching algorithm with respect to the pattern length.

Let $\Sigma$ denote an alphabet, i.e. a set of *letters* or *characters*, of cardinality $\sigma$. We assume a lexicographic order $<$ on $\Sigma$, naturally extended to the set of all strings over $\Sigma$. Letters in a string are numbered from 1.

## 2 Evenly spaced sparse suffix tree

We consider *evenly spaced sparse suffix trees* as defined in [8]. Consider a string $T[1..n]$. Let $Suf_r$ be the set of suffixes $\{T[rj+1..]|j=0,1,\ldots,\frac{n}{r}-1\}$ (assume for simplicity that $n$ is a multiple of $r$). Indexes $j$ will be called *ordinals*. An ordinal $j$ identifies the boundary between positions $rj$ and $rj+1$ in $T$, and the corresponding suffix $T[rj+1..]$. An *$r$-spaced suffix tree* of $T$, denoted $ST_r$, is a compacted trie for the set $Suf_r$. For $r=1$, the $r$-spaced suffix tree is the usual suffix tree. Similarly to the regular suffix trees, edges of an *$r$-spaced suffix tree* are labeled by substrings $T[i..j]$ of $T$, represented by a pair $(i,j)$. We define *explicit* and *implicit* nodes of $ST_r$ in the same way as for the regular suffix trees. Like in the regular suffix tree, an implicit node will be specified by a pair $(v,\ell)$, where $v$ is the closest explicit ancestor node and $\ell$ is the offset with reference to $v$. Note that by definition of the tree, the labels of the outgoing edges of any explicit node have different first letters.

Assuming that the last letter of $T$ is unique, $ST_r$ has $\frac{n}{r}$ leaves and then no more than $\frac{n}{r}$ explicit internal nodes. Therefore, $ST_r$ takes $O(\frac{n}{r})$ space.

By default, a *node* may refer to either an explicit or an implicit node. A string $\alpha$ is *represented* in $ST_r$ if $\alpha$ is a prefix of one of the suffixes of $Suf_r$, i.e. if $\alpha$ is a substring of $T$ starting at a position $rj+1$ for some $j$. In this case, $\alpha$ is the label of some node $v$ of $ST_r$, and we say that $\alpha$ *is represented by $v$*, and $|\alpha|$ is the *string depth* of $v$.

Similarly to $r$-spaced suffix trees, we define an $r$-spaced suffix array. Consider the lexicographic order on suffixes $Suf_r$ and define $SA_r[i] = j$ iff $i$ is the rank of $T[rj+1..]$ in the lexicographic order on $Suf_r$. Since $SA_r$ is a permutation of the ordinals $\{0,1,\ldots,\frac{n}{r}-1\}$, there is an inverse mapping, denoted $SA_r^{-1}$. Thus, for the suffix of $T$ starting at a position $rj+1$ of $T$, its number in lexicographic order on $Suf_r$ equals $SA_r^{-1}[j]$.

Note that each leaf of the tree $ST_r$ represents some suffix of $Suf_r$, and we call the *rank* of a leaf $v$ the rank of the suffix represented by $v$ in the lexicographic order on $Suf_r$. Note that the rank of $v$ is equal to $SA_r^{-1}[j]$, where $T[rj+1..]$ is the suffix of $T$ represented by $v$.

If the children of each internal node of $ST_r$ are ordered by the lexicographic order of the labels of corresponding edges, then the leaves of $ST_r$ (as occurring in the depth-first traversal) become ordered by their ranks. For a node $v$, we define $MinRank(v)$ and $MaxRank(v)$ to be respectively the minimal and the maximal rank of leaves in a subtree of $ST_r$ rooted at $v$. The ranks of all leaves of the subtree rooted at $v$ form the *rank interval* $[MinRank(v), MaxRank(v)]$.

If $\alpha$ is a word corresponding to $v$, then the ranks of suffixes of $Suf_r$ starting with $\alpha$ are specified by the interval $[MinRank(v), MaxRank(v)]$.

We assume that for each explicit node $v$ of $ST_r$, $MinRank(v)$ and $MaxRank(v)$, as well as its string depth $d(v)$ can be recovered in constant time. This can be trivially achieved by post-processing the tree and storing this information explicitly.

We extend the $r$-spaced suffix tree $ST_r$ with *suffix links* defined differently than in [8]. For each explicit node $v$ representing a string $\alpha$, a suffix link $s(v)$ maps $v$ to a (not necessarily explicit) node labeled with the longest proper suffix $\alpha[i+1..]$ of $\alpha$ represented in the tree.

Offset $i$ will be called the *type* of the suffix link. It follows easily that $1 \leq i \leq r$. For each explicit node $v$ of $ST_r$, we store the target node $s(v)$ together with the type of the suffix link.

Given a string $T$, the $r$-spaced suffix tree $ST_r$ including functions $s$, $SA_r$ and $SA_r^{-1}$ can be constructed in time $O(n)$ and space $O(\frac{n}{r})$. Due to space limitations, the construction is described in Appendix B.

## 3 RIGHTSEARCH

Consider a pattern $P[1..m]$. Using the sparse suffix tree, we locate all occurrences of pattern suffixes $P[1..], P[2..], \ldots, P[r..]$ at block boundaries using a procedure RIGHTSEARCH that we describe in this section. Once an occurrence of $P[k+1..]$ is found, the rank interval of $P[k+1..]$ in $SA_r$ is submitted to another procedure LEFTSEARCH that selects from it the positions which are preceded by occurrences of $P[1..k]$, thus locating the whole pattern. We will say in this case that $P$ occurs in $T$ with $k$-offset. LEFTSEARCH will be described in Sections 4-5.

RIGHTSEARCH proceeds by navigating through $ST_r$ trying to locate all nodes representing $P[1..], P[2..], \ldots, P[r..]$. Starting at the root with $P[1..]$, RIGHTSEARCH follows down the current suffix $P[k+1..]$ in the tree as long as possible.

When following an edge in the tree, its label $T[i..j]$ is divided into blocks of $r$ letters and each block, except possibly the last incomplete block, is compared by a single operation. The last incomplete block is compared letter-by-letter.

The pseudo-code of RIGHTSEARCH is given in Algorithm 1 in Appendix A. Assume that RIGHTSEARCH arrives at some (generally implicit) node $(v, \ell)$ reaching the end of $P[k+1..m]$ (line 14 of Algorithm 1). Then the algorithm retrieves the rank interval $[MinRank(v'), MaxRank(v')]$, where $v'$ is the closest explicit descendant node, which specifies all the occurrences of $P[k+1..]$ at block boundaries. After that, the traversal jumps to $s(v)$ and proceeds with the prefix $P[k+i+1, m-\ell+1]$ of the current suffix $P[k+i+1..]$, where $i$ is the type of suffix link $s(v)$ (lines 20-22).

Assume now that RIGHTSEARCH riches a mismatch while processing current suffix $P[k+1..]$ (line 8). Assume that the mismatch occurred when visiting a node $(v, \ell)$ and processing a prefix $P[k+1..p]$ of $P[k+1..]$. Similarly to the previous case, the algorithm jumps to $s(v)$ and proceeds with the prefix $P[k+i+1, p-\ell+1]$ of the new current suffix $P[k+i+1..]$, where $i$ is the type of suffix link $s(v)$.

Importantly, the described procedure does not miss any occurrences:

**Lemma 1.** *Algorithm 1 correctly identifies all suffixes $P[k+1..]$, $0 \leq k \leq r-1$, occurring at block boundaries of $T$.*

*Proof.* It is easy to see by induction that once a suffix $P[k+1..]$ is found (line 11 of Algorithm 1), it is represented in the tree and therefore occurs starting at a block boundary.

A key point is that the procedure does not miss any such suffixes. This is due to the definition of suffix links: when following a suffix link (lines 20-22), the algorithm switches from processing the suffix $P[k+1..]$ to the suffix $P[k+i+1..]$, where $i$ is the type of the suffix link. It follows that no suffix $P[k+i'+1..]$ for $i' < i$ can be represented in the tree. This is because the suffix link points to the *longest* suffix represented in the tree. □

Let us now turn to the analysis of the running time of RIGHTSEARCH. The algorithm navigates over the suffix tree $ST_r$ by following edges downwards, either by chunks of $r$ letters or letter-by-letter, and by following suffix links. We analyse separately the traversal of two types of edges: completely traversed edges (hereafter *traversed edges*), and incompletely traversed edges (hereafter *dead-end edges*), either due to a mismatch or due to a found suffix.

The number of dead-end edges is at most $r$, as each of them terminates the processing of some suffix $P[k+1..]$. On each such edge, the algorithms makes no more that $m/r$ block comparisons and $r$ letter-by-letter comparisons. Therefore, the whole time spent on dead-end edges is $O(m + r^2)$.

The number of all comparisons made along the traversed edges is bounded by $m$, as these comparisons compare different portions of the pattern. In other words, the sequence of these comparisons can be associated with moving a pointer in the pattern left-to-right, either by blocks of $r$ letters or by single letters. The whole time spent on these comparisons is thus $O(m)$.

**Theorem 1.** RIGHTSEARCH *computes the rank intervals of all suffixes $P[k+1..]$, $0 \leq k \leq r-1$, occurring at block boundaries of $T$ in time $O(m + r^2)$.*

## 4  Compacted trie

RIGHTSEARCH, described in the previous section, computes rank intervals of all pattern suffixes $P[k+1..]$ occurring at block boundaries of $T$. For each such suffix, procedure LEFTSEARCH is called, which selects, from this rank interval of $SA$, those boundary positions which are preceded by $P[1..k]$. LEFTSEARCH is based on another data structure – compacted trie of reversed blocks – which we describe in this section.

The data structure, denoted $CT_r$, is based on a compact trie storing all the blocks of $T$ written in reverse order, i.e. all the strings $\tau_j = (T[r(j-1)+1..rj])^R$ for $j = 1, \ldots, \frac{n}{r}$. Since there are $\frac{n}{r}$ blocks overall, $CT_r$ has no more than $\frac{n}{r}$ leaves and therefore takes $O(\frac{n}{r})$ space.

For each node $v$ of $CT_r$, let $l(v)$ be the string represented by $v$ in $CT_r$, and let $d(v) = |l(v)|$ be the *string depth* of $v$.

$CT_r$ is used by LEFTSEARCH in a natural way: in order to find occurrences of $P[1..k]$ ending at block boundaries, we look up its reverse $P[k]P[k-1]..P[1]$ in $CT_r$ by following the corresponding branch. However, selecting efficiently those occurrences which belong to the rank interval computed by RIGHTSEARCH is a non-trivial task which can be reduced to some kind of orthogonal range queries problem (see [3, 7]). We propose a more efficient direct solution inspired by the technique used in [10] for the problem of $3D$-dominance reporting. The technique is also somewhat similar to wavelet trees [5] which have become a popular tool in text compression and indexing (see [9]).

For each node $v$ of $CT_r$, let $Ord_v = <j_1, j_2, \ldots, j_{N(v)}>$ be an ordered set of all ordinals $j$ such that $l(v)$ is a prefix of $\tau_j$, that is $(l(v))^R$ occurs in $T$ ending at a position $rj$. The order of ordinals $j_1, j_2, \ldots, j_{N(v)}$ is defined by their rank in the lexicographic order on $T[rj+1..]$ (see Section 2). In other words,

$$SA^{-1}[j_1] < SA^{-1}[j_2] < \ldots < SA^{-1}[j_{N(v)}].$$

$Ord_v$ is not stored explicitly for internal nodes of $CT_r$ but is stored explicitly for the leaves. For each leaf $v$ of $CT_r$, $Ord_v$ is stored as an array of $N(v)$ entries containing $j_1, j_2, \ldots, j_{N(v)}$ in order. For each internal node $v$, we store two arrays, $\rho_v$ and $c_v$, that we describe now.

The array $\rho_v$ contains letters stored in packed form, i.e. each entry of $\rho_v$ is a machine word that stores some fixed number of letters to be defined later. The letter sequence stored in $\rho_v$ is defined as follows. If $Ord_v = <j_1, \ldots, j_{N(v)}>$, then $\rho_v$ contains letters $T[rj_1 - d(v)], T[rj_2 - d(v)], \ldots, T[rj_{N(v)} - d(v)]$ in this order.

Array $\rho_v$ provides information necessary for choosing an appropriate child when navigating down through $CT_r$. If $u_1, u_2, \ldots, u_t$ are children of $v$, then $Ord_v = Ord_{u_1} \uplus Ord_{u_2} \uplus \ldots \uplus Ord_{u_t}$. Consider $j \in Ord_v$. Then $j \in Ord_{u_i}$ iff $T[rj - d(v)]$ is the first letter of the label of the edge $(v, u_t)$.

We now define the size of $\rho_v$ which is determined by the number of letters stored in one entry. Our basic assumption is that machine word is at least $\log n$ bits, and therefore can hold at least $\frac{\log n}{\log \sigma} = \log_\sigma n$ letters. Then, assuming $r \leq \log_\sigma n$ insures that a block of $r$ letters can be compared with unit cost, which is our primary condition (see Introduction). Define the number of letters to be stored in one entry of $\rho_v$ to be $r/2$. The size of $\rho_v$ is then $\frac{2N(v)}{r}$ machine words.

The second array $c_v$ is a 2-dimensional array of integers. For each letter $b \in \Sigma$ and each $j = 1, 2, \ldots, \frac{2N(v)}{r}$, $c_v[b, j]$ stores the number of occurrences of letter $b$ in the first $j$ machine words of $\rho_v$. It takes $\frac{2\sigma N(v)}{r}$ memory words to store $c_v$.

The number of letters inside each entry of $\rho_v$ is preprocessed separately, following the Four-Russians idea. Formally, on top of the data structure $CT_r$ described above, we store a 3-dimensional array $C$ defined as follows. For each possible instance $u$ of entry of $\rho_v$, each letter $b \in \Sigma$ and each $j = 1, 2, \ldots, r/2$, $C[u, b, j]$ stores the number of letters $b$ among the first $j$ letters contained in $u$.

**Lemma 2.** *The array $C$ takes $o(\frac{n}{r})$ space.*

*Proof.* As each entry of $\rho_v$ contains $r/2$ letters, the number of possible entries is $\sigma^{r/2}$. Therefore, the size of $C$ is $\sigma^{r/2+1}(r/2)$. As $r \leq \log_\sigma n$, the size of $C$ is $O(n^{1/2} \log n) = o(\frac{n}{r})$. $\square$

The following lemma summarizes the space taken by all the data structures.

**Lemma 3.** *Data structure $CT_r$ and the array $C$ take space $O(\frac{n}{r})$ altogether.*

*Proof.* The compact trie $CT_r$ has $\frac{n}{r}$ leaves and then no more than $\frac{n}{r}$ internal nodes. The ordinal sets stored at leaves are pairwise disjoint and hold all the ordinals $j = 1, \ldots, \frac{n}{r}$ altogether, therefore their total size is $\frac{n}{r}$.

Each letter of $T$ appears at most once in all arrays $\rho_v$, which implies that the overall number of letters in all $\rho_v$ is $O(n)$. Therefore, the overall size of all $\rho_v$ is $O(\frac{n}{r})$. For the same reason, the overall size of arrays $c_v$ is $O(\frac{n}{r})$ too. The size of $C$ is $o(\frac{n}{r})$. The Lemma follows. $\square$

In Appendix C we will show how to construct $CT_r$ including all the arrays $\rho_v$, $c_v$ and $C$, in $O(n)$ time and $O(\frac{n}{r})$ space.

## 5 LEFTSEARCH

We now describe the algorithm LEFTSEARCH. Recall that RIGHTSEARCH locates all suffixes $P[k+1..]$, $0 \leq k \leq r-1$ occurring at block boundaries. For each such suffix $P[k+1..]$, RIGHTSEARCH outputs the corresponding rank interval $[LB^k, RB^k]$ in $SA_r$ such that $P[k+1..]$ occurs precisely at positions $\{rj + 1 | j \in SA_r[i], LB^k \leq i \leq RB^k\}$. The goal of LEFTSEARCH is to select from this set those positions which are preceded by the prefix $P[1..k]$.

Let us fix some $k$, $1 \leq k \leq r-1$. (For $k = 0$, RIGHTSEARCH finds the entire occurrence of $P$ and therefore LEFTSEARCH is not needed.) As already mentioned in Section 4, the general idea of LEFTSEARCH is intuitive: the algorithm simply follows $P[k]P[k-1]\ldots P[1]$ in $CT_r$ starting from the root. If this word is not represented in $CT_r$, then $P[1..k]$ does not have any occurrences ending at block boundaries.

Assume that $v_0, v_1, \ldots, v_\ell$ are nodes of $CT_r$ traversed when following $P[k]P[k-1]\ldots P[1]$. Consider a node $v_i$ of string depth $d(v_i)$ and the associated ordered set $Ord_{v_i}$. The following statement holds.

**Lemma 4.** *The ordinals $j$ such that $P[k - d(v_i) + 1..k]$ is a suffix of $T[..rj]$ and $P[k+1..]$ is a prefix of $T[rj + 1..]$ form an interval of $Ord_{v_i}$.*

*Proof.* $Ord_{v_i}$ is, by definition, the set of ordinals $j$ such that $P[k - d(v_i) + 1..k]$ is a suffix of $T[r(j-1) + 1..rj]$. These ordinals $j$ are ordered in $Ord_{v_i}$ according to the lexicographic ordering of suffixes $T[rj + 1..]$ of $T$. Those suffixes which start with $P[k+1..]$ form then an interval of $Ord_{v_i}$. $\square$

Lemma 4 provides the key idea of LEFTSEARCH: when following $P[k]P[k-1]\ldots P[1]$ in $CT_r$ for each visited node $v_i$ maintain the interval of $Ord_{v_i}$, which contains those ordinals $j$ of $Ord_{v_i}$ for which $P[k+1..]$ is a prefix of $T[rj+1..]$. Note that since sets $Ord_{v_i}$ are not stored explicitly for internal nodes, the algorithms actually manipulates indexes $1,\ldots,N(v_i)$ of ordinals $Ord_{v_i}$ rather than ordinals themselves. More precisely, when visiting a node $v_i$, the algorithm will compute the corresponding interval $[LB(v_i), RB(v_i)]$ of $[1..N(v_i)]$.

The traversal of $CT_r$ by word $P[k]P[k-1]\ldots P[1]$ starts at the root node $v_0$. The set $Ord_{v_0}$ is the set of all ordinals $\{0,\ldots,\frac{n}{r}-1\}$ ordered by the lexicographic order of suffixes $P[rj+1..]$, that is $Ord_{v_0} = < SA[i] | i = 1..\frac{n}{r} >$. The initial interval $[LB(v_0), RB(v_0)]$ of $Ord_{v_0}$ should contain those ordinals $j$ for which $P[k+1..]$ is a prefix of $T[rj+1..]$, which is precisely the rank interval $[LB^k, RB^k]$ computed by RIGHTSEARCH (see Algorithm 1).

Let us now focus on the key step of LEFTSEARCH which consists in updating the current interval $[LB(v_i), RB(v_i)]$ when moving in $CT_r$ from a current node $v_i$ to one of its children.

Let $v_i$ be an internal node of $CT_r$ reached by word $P[k]\ldots P[k-d(v_i)+1]$, and $[LB(v_i), RB(v_i)]$ be the interval of $Ord_{v_i}$ computed by the algorithm. Consider the next letter $a = P[k-d(v_i)]$, and the child $v_{i+1}$ reached by $P[k]\ldots P[k-d(v_i)+1]P[k-d(v_i)]\ldots P[k-d(v_{i+1})+1]$. Let $a$ be the first letter of the label of edge $(v_i, v_{i+1})$, i.e. $a = P[k-d(v_i)]$. The following lemma shows how to compute the interval $[LB(v_{i+1}), RB(v_{i+1})]$ from the interval $[LB(v_i), RB(v_i)]$.

**Lemma 5.** $LB(v_{i+1}), RB(v_{i+1})$ can be obtained from $LB(v_i), RB(v_i)$ by the following formulas:

$$LB(v_{i+1}) = c_{v_i}[a, \left\lfloor \frac{LB(v_i)}{r/2} \right\rfloor] + C[\rho_{v_i}[\left\lceil \frac{LB(v_i)}{r/2} \right\rceil], a, LB(v_i) \setminus (r/2)], \quad (1)$$

$$RB(v_{i+1}) = c_{v_i}[a, \left\lfloor \frac{RB(v_i)}{r/2} \right\rfloor] + C[\rho_{v_i})[\left\lceil \frac{RB(v_i)}{r/2} \right\rceil], a, RB(v_i) \setminus (r/2)], \quad (2)$$

where $\setminus$ denotes the remainder of integer division.

*Proof.* Let us first analyse how $Ord_{v_{i+1}}$ is related to $Ord_{v_i}$. It is easily seen that $Ord_{v_{i+1}} \subseteq Ord_{v_i}$ and the order of elements of $Ord_{v_{i+1}}$ is preserved in $Ord_{v_i}$. Furthermore, $j \in Ord_{v_{i+1}}$ iff $j \in Ord_{v_i}$ and $T[r(j-1) - d(v_i)] = P[k - d(v_i)]$. (Note that since there are no branching nodes between $v_i$ and $v_{i+1}$, this means that $T[r(j-1) - d(v_{i+1}) + 1..r(j-1) - d(v_i)] = P[k - d(v_{i+1}) + 1..k - d(v_i)]$ and then $T[r(j-1) - d(v_{i+1}) + 1..r(j-1)] = P[k - d(v_{i+1}) + 1..k]$.) Therefore, computing interval $[LB(v_{i+1}), RB(v_{i+1})]$ from $[LB(v_i), RB(v_i)]$ can be done through counting the number of $a$'s among all the letters stored in $\rho_{v_i}$ and within intervals of $\rho_{v_i}$ defined by positions $LB(v_i), RB(v_i)$. These counts can be retrieved from auxiliary arrays $c_{v_i}$ and $C$.

Recall from Section 4 that letters are stored in $\rho_{v_i}$ in packed form, by $r/2$ letters in each machine word. The number of $a$'s within several consecutive machine words is provided by arrays $c_{v_i}$, whereas the number of $a$'s within a part of one machine word is provided by the array $C$. Formulas (1),(2) follow. The computation is illustrated in Figure 1. □

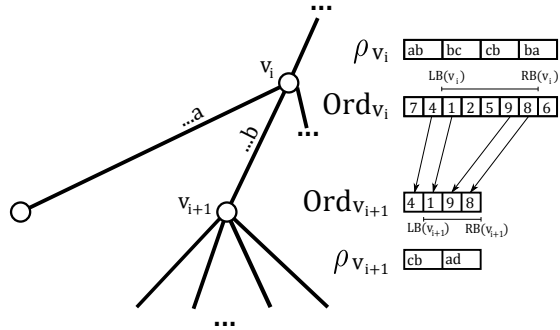

Figure 1: Illustration to the key step of LeftSearch (Lemma 5). In this example, $r = 4$ and each machine word of $\rho_v$ stores 2 letters. $[LB(v_i), RB(v_i)] = [2, 7]$. The first letter of the label on the edge $(v_i, v_{i+1})$ is $b$. Therefore, $LB(v_{i+1}) = 1$ as there is one letter $b$ among the first two letters of $\rho_{v_i}$. $RB(v_{i+1})$ is computed similarly. Arrays $Ord_v$ are not stored for internal nodes and are shown here for explanatory purposes only.

Algorithm 2 in Appendix A shows the pseudo-code of LeftSearch. Based on Lemma 5, LeftSearch recomputes, using formulas (1),(2), the current interval $[LB(v_i), RB(v_i)]$ at each node $v_i$ along the traversal of the branch of $CT_r$ defined by word $P[k] \ldots P[1]$. If at some point the interval $[LB(v_i), RB(v_i)]$ gets empty, this implies that there is no occurrence of $P$ with $k$-offset, and LeftSearch terminates. Once the terminal node $v_\ell$ is reached[1], the algorithm has identified a subtree of $CT_r$ such that the leaves of this subtree store the ordinals $j$ corresponding to the occurrences of $P[1..k]$ ending at block boundaries. For each such leaf $u$, the set $Ord_u$ of these ordinals is explicitly stored in an array. However, similar to internal nodes, for each leaf $u$ we have to compute the interval $[LB(u), RB(u)]$ defining the ordinals of interest. This is done in the same way as before, namely by traversing down the branches of $CT_r$ and updating the interval using formulas (1),(2). The only difference is that starting from $v_\ell$ we need to explore *all branches* of $CT_r$, rather than only one branch determined by the word $P[k]..P[1]$.

Thus, the algorithm proceeds with exploring all the branches of the subtree defined by $v_\ell$ and performing the computation of Lemma 5 for *all* the children of each node, rather than for only one child as before. An obvious optimization here is that once a current interval $[LB(v), RB(v)]$ gets empty for some node $v$, the algorithm stops exploring the subtree of $v$, as none of its leaves can contain the desired ordinals. The traversal of the subtree of $v_\ell$ is done by an auxiliary

---
[1] The node reached after following $T[k]..T[1]$ can be an implicit node of $CT_r$, in which case the algorithm moves on to the closest explicit descendant node (lines 16-18 in Algorithm 2)

procedure TRAVERSE shown in Algorithm 3 in Appendix.

We are left with the analysis of the running time of LEFTSEARCH. Since the computation of Lemma 5 is done in constant time, the traversal of $v_0, v_1, \ldots, v_\ell$ is done in time $O(k)$, that is $O(r)$. Starting from $v_\ell$, the algorithm explores the corresponding subtree but once the current interval $[LB(v), RB(v)]$ gets empty, the subtree of $v$ is pruned out.

Let us call a node $v$ (internal or a leaf) *non-empty* if the corresponding interval $[LB(v), RB(v)]$ is non-empty. Observe that for a non-empty internal node at least one of its descendants is non-empty. This means that there is no non-empty internal nodes outside the paths leading to non-empty leaves. Processing every non-empty internal node requires $O(\sigma)$ time, which is the time to examine its ancestors. The whole traversal of the subtree of $v_\ell$ takes time $O(\sigma r)$ per non-empty leaf.

Since every non-empty leaf defines at least one $k$-offset occurrence of $P$, the total running time of LEFTSEARCH is then $O(r + r \cdot occ_k)$, where $occ_k$ is the number of resulting $k$-offset occurrences.

## 6 Resulting bound

**Theorem 2.** *Searching for all occurrences of $P$ in $T$ using algorithms RIGHTSEARCH and LEFTSEARCH takes time $O(m + r^2 + r \cdot occ)$, where $occ$ is the total number of output occurrences.*

*Proof.* Time taken by RIGHTSEARCH is $O(m+r^2)$. There are at most $r$ calls to LEFTSEARCH that take time $O(r \cdot r + r \sum_k occ_k) = O(r^2 + r \cdot occ)$. The theorem follows. □

Note for completeness that we have always assumed that the pattern length $m$ is larger than $r$ and therefore must cross at least one block boundary. In case $m < r$, all occurrences of $P$ located inside blocks can be reported in time $O(m + occ)$ (see [3]).

## 7 Concluding remarks

In this paper, we proposed a compact indexing scheme supporting linear-time string matching. The guiding idea is the packing of several characters into one machine word and the use of the sparse suffix tree based on partitioning the text string into blocks of equal size $r$. The core of the algorithm is the procedure RIGHTSEARCH computing, in a single traversal of the sparse suffix tree, all the suffixes $P[k+1..]$, $0 \leq k < r$, in time $O(m + r^2)$. For each such suffix, procedure LEFTSEARCH is called which selects those occurrences of $P[k+1..]$ which are preceded by $P[1..k]$. All resulting occurrences are then reported in time $O(m + r^2 + r \cdot occ)$.

One of our goals was to design a simple dedicated algorithm that does not call for complex external subroutines, such as the one supporting orthogonal range queries. The obtained solution, however, remains somewhat complex: in particular, the additional compact trie data structure and the implementation of LeftSearch represent a complex step. We believe that this could be further improved leading to a simpler algorithm possibly using only one data structure. Such a simplification could possibly lead to getting rid of the $r$ factor in the $r \cdot occ$ term of the complexity bound, thus yielding a fully linear solution both on the pattern length and the number of pattern occurrences. This constitutes a challenging problem for future research.

# References

[1] A. Andersson, J. Larsson, and K. Swanson. Suffix trees on words. In *Proc. of the 7th Annual Symposium on Combinatorial Pattern Matching (CPM'96)*, volume 1075 of *Lecture Notes in Computer Science*, pages 102–115. Springer, 1996.

[2] Ph. Bille. Fast searching in packed strings. In *Proc. of the 20th Annual Symposium on Combinatorial Pattern Matching (CPM'09), Lille, France, June 22-24, 2009*, volume 5577 of *Lecture Notes in Computer Science*, pages 116–126. Springer, 2009.

[3] Y.-F. Chien, W.-H. Hon, R. Shah, and J. Vitter. Geometric burrows-wheeler transform: Linking range searching and text indexing. In *Proc. Data Compression Conference (DCC 2008)*, pages 252–261. IEEE Computer Society, 2008.

[4] S.-Y. Chiu, W.-K Hon, R. Shah, and J. Vitter. I/O-efficient compressed text indexes: From theory to practice. In *Proc. Data Compression Conference (DCC 2010)*, pages 426–434. IEEE Computer Society, 2010.

[5] R. Grossi, A. Gupta, and J. Vitter. High-order entropy-compressed text indexes. In *Proc. 14th Annual ACM-SIAM Symposium on Discrete Algorithms (SODA)*, pages 841–850. SIAM, 2003.

[6] J. Healy, E. Thomas, and J. Schwartz. Annotating large genomes with exact word matches. *Genome Reseach*, 13:2306–15, 2003.

[7] W.-K. Hon, R. Shah, Sh. Thankachan, and V. Vitter. On entropy-compressed text indexing in external memory. In *Proc. 16th Int. Symp. String Processing and Information Retrieval (SPIRE'09)*, volume 5721 of *Lecture Notes in Computer Science*, pages 75–89. Springer, 2009.

[8] J. Kärkkäinen and E. Ukkonen. Sparse suffix trees. In *Proc. 2nd Annual International Computing and Combinatorics Conference (COCOON'96)*, volume 1090 of *Lecture Notes in Computer Science*, pages 219–230. Springer Verlag, 1996.

[9] V. Mäkinen and G. Navarro. Compressed full-text indexes. *ACM Computing Surveys*, 39(1), 2007. Article 2.

[10] Q. Shi and J. JáJá. Fast algorithms for 3-d dominance reporting and counting. *Int. J. Found. Comput. Sci.*, 15(4):673–684, 2004.

[11] J. Simpson and R. Durbin. Efficient construction of an assembly string graph using the FM-index. *Bioinformatics*, 26(12):i367–i373, 2010.

# A  Pseudocodes

---
**Algorithm 1** RIGHTSEARCH
---
1: $k \leftarrow 1$
2: $p \leftarrow 1$
3: $Vertex \leftarrow root$
4: $VertexOffset \leftarrow 0$

5: **while** $k < r$ **do**
6:    **while** $p \leq m$ **do**
7:       follow down the current edge of $ST_r$ by comparing at once $r$ characters $P[p..p+r-1]$ if possible, or one character $P[p]$ otherwise
8:       **if** mismatch occurred **then**
9:          break the **while**-loop
10:       **else**
11:          update $Vertex, VertexOffset, p$
12:       **end if**
13:    **end while**

14:    **if** $p = m$ **then**
15:       **if** $VertexOffset \neq 0$ **then**
16:          $Descendant \leftarrow$ closest explicit descendant for $(Vertex, VertexOffset)$
17:       **end if**
18:       LEFTSEARCH$(k, MinRank(Descendant), MaxRank(Descendant))$
19:    **end if**

20:    $p \leftarrow p - VertexOffset + 1$
21:    $(Vertex, VertexOffset) \leftarrow s(Vertex)$
22:    $k \leftarrow k +$ type of the suffix link $(Vertex, s(Vertex))$
23: **end while**

**Algorithm 2** LeftSearch($k, L, R$)

1: $CurrVertex \leftarrow root$
2: $LB \leftarrow L$
3: $RB \leftarrow R$
4: $i \leftarrow k$
5: **while** $i \geq 1$ **and** $[LB, RB]$ is not empty **do**
6:   **if** $P[i]$ matches an outgoing label **then**
7:     move down by $P[i]$
8:     update $CurrVertex$
9:     update $LB, RB$ using formulas (1),(2)
10:     $i \leftarrow i - 1$
11:   **else**
12:     **return** no occurrences
13:   **end if**
14: **end while**

15: **if** $[LB, RB]$ is not empty **then**
16:   **if** $CurrVertex$ is implicit **then**
17:     $CurrVertex \leftarrow$ closest explicit descendant of $CurrVertex$
18:   **end if**
19:   Traverse ($CurrVertex, LB, RB$)
20: **end if**

---

**Algorithm 3** Traverse($Vertex, LB, RB$)

1: **for all** sons $u$ of $Vertex$ **do**
2:   Compute $LB(u), RB(u)$ using formulas (1),(2)
3:   **if** $[LB(u), RB(u)]$ is not empty **and** $u$ is not a leaf **then**
4:     Traverse($u, LB(u), RB(u)$)
5:   **else if** $u$ is a leaf **then**
6:     **for all** $i \in [LB(u), RB(u)]$ **do**
7:       output $Ord_u[i]$
8:     **end for**
9:   **end if**
10: **end for**

# B  Construction of $ST_r$

In this section we describe the construction of the suffix tree $ST_r$, as defined in Section 2, for a given text string $T$.

Algorithms to construct the sparse suffix tree in time $O(n)$ and space $O(n/r)$ have been proposed in [8] (see also [1]). However, the definition of sparse suffix tree from [8] differs from ours in the definition of suffix links. Specifically, according to [8], a suffix link from an explicit node representing a string $\alpha$ points to a node representing $\alpha[r+1..]$. We call such suffix links $r$-suffix links. The definition is well-founded, as implied by the following lemma:

**Lemma 6.** *If a string $\alpha$, $|\alpha| > r$, is represented in $ST_r$, then the string $\alpha[r+1..]$ is represented in $ST_r$ too. Moreover, if $\alpha$ is represented by an explicit node, then the same holds for $\alpha[r+1..]$.*

Assume that the sparse suffix tree together with $r$-suffix links has already been constructed by the algorithm of [8]. To obtain $ST_r$, we have only to set the suffix links as defined in Section 2. We will be setting suffix links consecutively for type $1, 2, \ldots, r$.

For each explicit node $v$ of $ST_r$, we fix one of the occurrences of the represented string $l(v)$ in $T$ starting at a block boundary. We then compile an array $Q$ of $\frac{n}{r}$ lists of nodes of $ST_r$. A node $v$ belongs to the $i$th list iff the fixed occurrence of $l(v)$ starts at position $ir + 1$ of $T$. We assume that nodes in each list of $Q$ occur in the increasing order of string depths. $Q$ can be compiled by one breadth-first traversal of $ST_r$ in $O(\frac{n}{r})$ time.

Consider some $i$, $0 \leq i \leq r-1$. Let $\beta_i^j$, $0 \leq j \leq \frac{n}{r} - 1$ be the longest prefix of $T[rj+i+1..]$ represented in $ST_r$. At the *first step* of the construction, the algorithm locates the (possibly implicit) nodes $v_0, v_1, \ldots, v_{\frac{n}{r}-1}$ of $ST_r$ representing $\beta_i^0, \beta_i^1, \ldots, \beta_i^{r-1}$ respectively. These nodes are used to build suffix links of type $i$.

The following lemma can be proved:

**Lemma 7.** *The nodes $v_0, v_1, \ldots, v_{\frac{n}{r}-1}$ can be located in time $O(\frac{n}{r})$.*

We leave the details of the proof for an extended version of the paper.

The *second step* is to build suffix links of type $i$ using the nodes $v_0, v_1, \ldots, v_{\frac{n}{r}-1}$.

**Lemma 8.** *Let $u$ and $v$ be two explicit nodes such that $u$ is an ancestor of $v$ (that is, $l(u)$ is a prefix of $l(v)$). Then the type of the suffix link of $u$ is not larger than the type of the suffix link of $v$.*

The Lemma will insure that all nodes with suffix links of type $i$ occur consecutively in the initial part of lists $Q[j]$ (note that by induction, the nodes with suffix links of type smaller than $i$ have been deleted from lists $Q[j]$, see below) and if the head element of some $Q[j]$ does not have a suffix link of type $i$, then no other element of $Q[j]$ has one. Note also that a suffix link of type $i$ of some node $v$ in $Q[j]$ must point to a node on the path from the root to $v_j$. Hence, the main idea is to maintain a stack of nodes on the path from the root of $ST_r$ to $v_j$ to compute suffix links of type $i$ for nodes of $Q[j]$. Note that $v_j$'s are implicit nodes in general, therefore some additional care is needed for this procedure.

In more details, we traverse $ST_r$ depth-first and maintain a stack $V$ (implemented as an array, i.e. allowing access to all stored elements) of size $O(\frac{n}{r})$ storing explicit nodes on the path from the root to the the current node of $ST_r$. Assume that we are in a node $v_j$ representing $\beta_i^j$, $0 \le j \le \frac{n}{r} - 1$. We check the head element $v$ of the list $Q[j]$. If the string depth $d(v)$ is less than $d(v_j)$, then the type of a suffix link from $v$ is $i$. We find the first node $u$ on the path from the root of $ST_r$ to $v_j$ with string depth bigger than $d(v)$ by a binary search on the elements of $V$. Obviously, the target node $s(v)$ is a (possibly implicit) node $(u, d(u) - d(v))$. After computing $s(v)$, $v$ is deleted from $Q[j]$. We repeat this procedure while string depth of the head element is less than $d(v_j)$ and then continue the tree traversal.

Let us now turn to time and space analysis. We need $O(n)$ time and $O(\frac{n}{r})$ space for construction of $ST_r$ and $SA_r$ for a string $T$ of length $n$. To compute $r$-suffix links we need $O(\frac{n}{r})$ time. To locate nodes $v_0, v_1, \ldots, v_{\frac{n}{r}-1}$ for a fixed $i$, we need $O(\frac{n}{r})$ time, and, therefore, $O(n)$ time for all $i, 0 \le i < r$. To compute all suffix links, we need $O(\frac{n}{r} \cdot \log \frac{n}{r} + n) = O(n)$ time. Finally, to store $V$ and $Q$ during tree traverses we need $O(\frac{n}{r})$ space.

## C  Construction of $CT_r$

In this section, we describe a construction algorithm for $CT_r$ (Section 4). First, note that the trie $CT_r$ for a string $T$ without additional arrays that we need can be constructed straightforwardly in time $O(n)$ and space $O(\frac{n}{r})$. Assume now that the trie has been constructed. We show how to augment it with arrays $c_v$, $\rho_v$ and $Ord_v$.

First, we compute the string depth for all nodes of the trie, which can be done in $O(\frac{n}{r})$ time by a depth-first traversal. The algorithm will proceed by depth levels, computing the auxiliary arrays for all nodes of depth 1, 2, etc. Note that the arrays $Ord_v$ are also stored explicitly for each level during the construction procedure, but are erased after processing the level (except for the leaves), for the sake of space economy. For each node of the current level, we store $Ord_v$ in lexicographical order and arrays $\rho_v$ and $c_v$ ($c_v$ is computed right after computing $\rho_v$ by one pass through $\rho_v$ in time $N(v)$).

It is enough to show that if we have computed arrays $c_v$, $\rho_v$ and $Ord_v$ for a node $v$, then we can compute these arrays for each of its children $u$. Consider $Ord_v = < j_1, j_2, \ldots, j_{N(v)} >$. By definition, a leave labeled with $\tau_{j_k}$ is in a subtree of $CT_r$ with the root $u$ such that the label of the edge from $(v, u)$ starts with letter $\rho_v[k]$. Therefore, we read $\rho_v$ and copy $j_k$ to $Ord_u$, where the first letter of the label on the edge from $v$ to $u$ is $\rho_v[k]$ (note that $u$ is unique). After that, we write the letter $\tau_{j_k}[d(u)+1]$ into $\rho_u$.

To finish, we delete $Ord_v$ and compute the array $c_u$. All in all, we spend $O(\frac{n}{r})$ time for computing arrays of each next level. Since there are no more than $r$ levels, we need $O(n)$ time for computing additional arrays for $CT_r$. Note that arrays $Ord_v$ for the leaves will be built automatically. Construction of the array $C$ in time $O(n)$ is trivial.